\input{aipcheck}

\documentclass[
    ,final            
  ]
  {aipproc}

\layoutstyle{6x9}

\begin{document}

\title{High Redshift Supermassive Black Holes: X--ray observations}

\author{Andrea Comastri}{
  address={INAF--Osservatorio Astronomico di Bologna, via Ranzani 1, 
I--40127 Bologna, Italy}
}
\author{the {\tt HELLAS2XMM} team}{address={INAF--Bologna, 
INAF--Arcetri, INAF--Roma, 
IASF--CNR--Milano, Universita' Roma 3, Italy}}
%

\begin{abstract}

The spectrum of the hard X--ray background records the history 
of accretion processes integrated over the cosmic time.
Several observational and theoretical evidences indicate that
a significant fraction of the energy density is obscured by 
large columns of gas and dust. 
X--ray surveys are the most efficient way to trace accretion onto
supermassive black holes, since obscured, accreting sources are 
more difficult to select at all other wavelengths. 
The current status of hard X--ray surveys 
after the recent observations carried out with {\it Chandra} and 
XMM--{\it Newton} satellites is reviewed 
along with the results of extensive 
follow--up multiwavelength observations.
In particular recent results concerning the 
physical and evolutive properties of the supermassive black holes
hosted by X--ray selected Active Galactic Nuclei 
at high redshifts will be  discussed.

\end{abstract}

\maketitle


\section{Introduction}

The advent of imaging instruments in
the 2--10 keV band, first onboard {\tt ASCA} and {\tt BeppoSAX} 
(\cite{Ueda}, \cite{Akiyama}, \cite{rdc}, \cite{ff99,ff01}, 
\cite{flf02}, \cite{giommi}) and then on
{\it Chandra} and XMM--{\it Newton}, 
has led to a dramatic step forward in the study of  
accretion-powered supermassive black holes hosted in the nuclei
of distant active galaxies (AGN).
Thanks to the superb capabilities of the {\it Chandra} observatory
fully exploited by means of deep exposures which have reached 
about 1 Msec in the {\it Chandra} Deep Field South
({\tt CDFS} \cite{gia02,ros02} and almost 2 Msec 
in the {\it Chandra} Deep Field North ({\tt CDFN}; \cite{davo03}), 
about 80--90\% of the 2--10 keV  X--ray Background (XRB,
see e.g. \cite{mushy}, \cite{gia01,gia02}, \cite{bra02}, \cite{has01}) 
has been resolved into discrete sources.  

The study of the hard X--ray source population is being pursued
combining deep sensitive {\it Chandra} observations with 
medium--deep (i.e. the Lockman Hole; \cite{has01})
and shallow large area (i.e. the {\tt HELLAS2XMM} serendipitous
survey; \cite{baldi}) XMM--{\it Newton} surveys, 
which nicely complement the high spatial resolution {\it Chandra} 
observations and allow to collect enough X--ray photons to perform 
spectral analysis.  

These studies confirm, at least qualitatively, the predictions of standard
AGN synthesis models for the XRB (e.g. \cite{sw89},
\cite{com95,com01}\cite{gil01}):
the high energy background radiation is mostly due to the
integrated contribution of obscured and unobscured AGN folded with the 
corresponding evolution of their luminosity function over the cosmic time.

According to these models, most (i.e. 70--80 \%)
of the X--ray light produced by accretion 
onto supermassive black holes is obscured by large columns of gas and dust.
Although obscured AGN are common in the local Universe (\cite{maiolino},
\cite{ris99})
and several examples of distant objects are being found in the
optical follow--up observations of hard X--ray surveys (\cite{nor02}
\cite{ste02}), 
the space density, cosmological evolution and absorption distribution 
of the energetically dominant component of the XRB is still 
subject to several uncertainties which prevent us from a better 
understanding of their physical nature and their role in models
of galaxy evolution. 
Since we know that basically all spheroidal galaxy in the nearby 
Universe contains a massive black hole \cite{geb00},  
the study of the evolutionary 
properties of active galaxies 
would provide key information on the assembling and feeding of 
supermassive black holes over the cosmic time.

Here we review the status of hard X--ray surveys and associated 
multiwavelength program and how they can help in solving 
some of the issues outlined above.

\section{The content of X--ray surveys}

Extensive campaigns of optical spectroscopy observations have 
been carried out to identify the counterparts of hard X--ray sources.
Redshifts could be obtained  for a few hundreds of objects 
in both the {\tt CDFN} and {\tt CDFS} reaching a spectroscopic 
completeness of 
the order of 50--60\%  down to a limiting magnitude of R $\simeq$ 24.
We refer to \cite{barg02} and \cite{szo03} for
a detailed description of the source breakdown.
The most important and somehow unexpected finding of the 
identification process concerns the optical appearence of 
hard X--ray sources. The large majority of them do not show broad 
optical and ultraviolet emission lines which are known to be 
common among the counterparts of soft (0.5--2 keV) {\tt ROSAT} sources
\cite{leh01}. The hard X--ray spectra and high X--ray luminosities 
strongly suggest that the  accreting black holes which 
power the X--ray emission are obscured by large columns of gas and dust
which estinguish also the optical--UV light.

The present spectroscopic completeness of the identifications in the 
{\tt CDFS} and {\tt CDFN} is such to allow the calculation 
of reliable redshift distributions and luminosity functions
and the comparison with the prediction of AGN synthesis models for the 
XRB.
Surprisingly enough, the observed redshift distribution shows 
a rather narrow peak in the range $z=0.7-1$ and is dominated by low 
luminosity objects (log$L_X$ = 42--44). This behaviour is significantly 
different from that observed by previous
shallower {\tt ROSAT} surveys and in contrast with the predictions of XRB
synthesis models from which a population dominated by high luminosity
AGN at $z=1.5-2$ was expected (\cite{com95},\cite{gil01}).  
Furthermore, evidence is
emerging (related to the difference above) of a luminosity dependence
in the number density evolution of both soft and hard X-ray selected
AGN (\cite{cowie03}, \cite{has03}).

Although the above described results are quite robust, they are limited
by the lack of redshift information for optically faint sources. 
Moreover, the small area covered,
which is of the order of 0.05--0.1 deg$^2$ in both {\tt CDFN} and {\tt 
CDFS}, makes deep {\it Chandra} surveys 
subject to the effects of field to field fluctuations (cosmic variance) 
and not well suited for the search of rare objects 
such as high luminosity, high redshift quasars. As an example 
there are only 6 AGN with logL$_{2-10keV} > 44$ and  z$>3$ 
in the {\tt CDFN} among the almost 300 already identified sources.

In order to build sizeable samples of high luminosity, high redshift
sources such as  highly obscured type 2 quasars, a much
wider area needs to be covered, of the order of a few square degrees.
The {\tt HELLAS2XMM} survey represents a first step towards
this objective being designed to cover, when completed, some 4 deg$^2$
above a 2--10 keV limiting flux of about 10$^{-14}$ erg cm$^{-2}$ s$^{-1}$.

The optical follow--up spectroscopic identification process is 
being carried out making use of the ESO telescopes (3.6m and VLT),
 the 3.5m Italian National Telescope Galileo (TNG), 
and is complemented by multiwavelength observations in the
near infrared (with {\tt ISAAC}) and radio (with the {\tt VLA} and
{\tt ATCA} telescopes) bands. We refer to \cite{bru03,ff03} for
a detailed discussion of the results of the identification 
process.

\section{Searching for high--z obscured sources}

Evidences for the presence of 
a population of highly obscured sources are obtained combining 
the X--ray and optical fluxes of the yet unidentified sources
in both deep and shallow surveys.

\begin{figure}
  \includegraphics[height=.65\textheight]{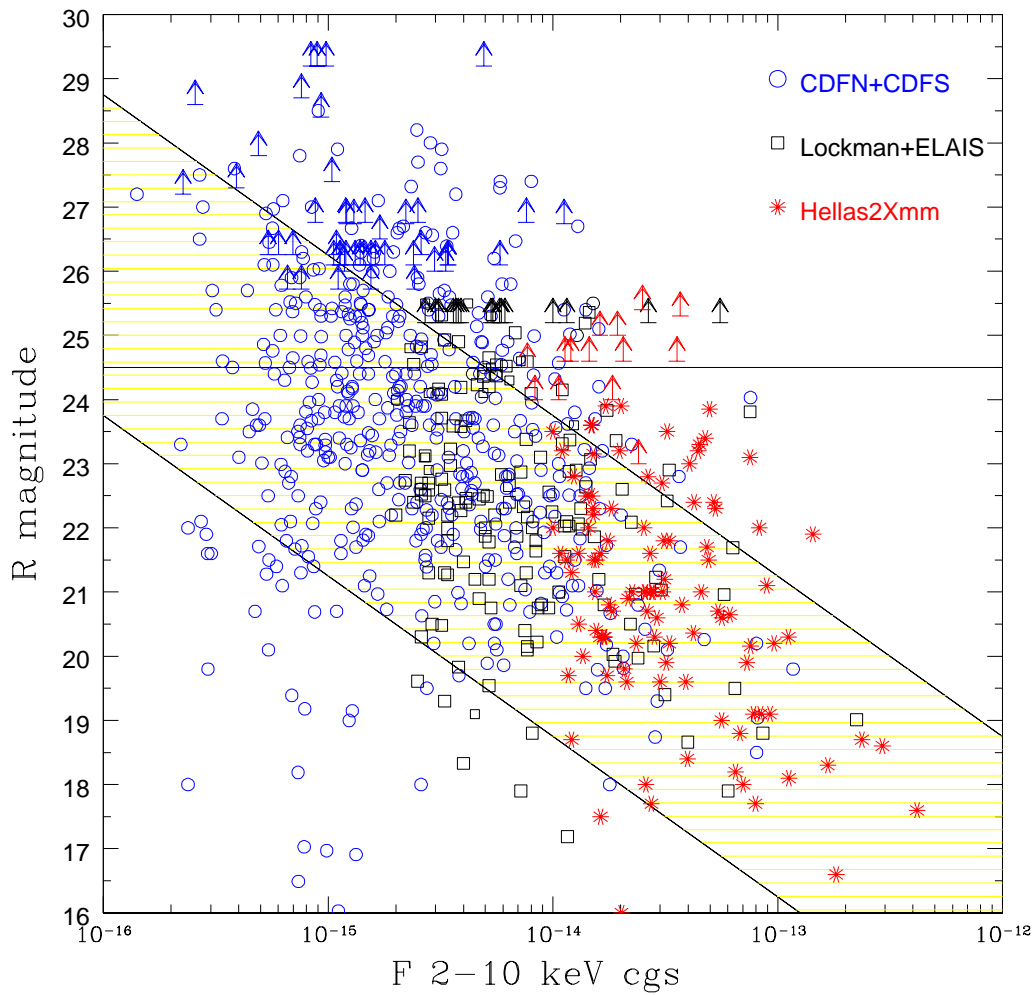}
  \caption{The 2--10 keV flux versus the optical R band magnitude for
several samples of hard X--ray selected sources. The part of the diagram below
the R=24.5 line is accessible to optical spectroscopy with 10m class 
telescopes. The shaded area comprised between the two diagonal lines, 
corresponding to log$f_X/f_{opt}$=1 (--1), represents the region 
occupied by ``conventional'' AGN (see \cite{cbm03})}
\end{figure}

While the large majority of spectroscopically identified broad line AGN 
fall within $-1 < log(f_X/f_{opt}) < $ 1 where $f_X$ is the 2--10 keV
X--ray flux and $f_{opt}$ the optical flux in the R filter, a sizeable 
fraction (of the order of 20--30 \%) of X--ray sources reported in 
Fig.~1 are characterized by extreme values of their $f_X/f_{opt}$ ratio. 
The ratio between the optical to X--ray optical depth, in the observer frame,
scales roughly as $(1+z)^{3.6}$, because dust extinction increases in
the UV while X--ray absorption strongly decreases going toward high
energies.  The net result is that in the presence of an absorbing
screen the observed optical flux of high--z QSO can be strongly
reduced, and the observed magnitudes can be mainly ascribed 
to the starlight of
the host galaxies. Conversely, the 2--10 keV X-ray flux can be much
less reduced.  Many extreme X-ray to optical flux ratio sources could 
therefore be distant, highly obscured type 2 QSO.

The spectroscopic identification of these objects
is already challenging the capabilities of ground based,  
10 m class optical telescopes calling for alternative, but less robust,
identification techniques, such as photometric redshifts or measures
based on the detection of redshifted iron lines.

It is interesting to note that the fraction of sources with high values
of $f_X/f_{opt}$ appears to be constant  over the 
entire range of X--ray fluxes sampled by deep and shallow surveys.
At the relatively bright fluxes covered by the {\tt HELLAS2XMM} survey
their optical counterparts are accessible to optical telescopes making
their spectroscopic identification possible. 
Indeed, among the 13  {\tt HELLAS2XMM}  sources 
with $f_X/f_{opt} > 10$ for which good quality VLT spectra are available,
we find 8 type 2 QSO classified as such on the basis of the lack of
of broad optical lines and high X--ray luminosity
(L$_{2-10keV}>10^{44}$ erg s$^{-1}$) \cite{ff03}.  

It seems reasonable to argue that most of the sources characterized 
by a high value of $f_X/f_{opt}$ are indeed high redshift
obscured AGN. Since high luminosity, obscured quasars 
are a key ingredient of XRB synthesis models and may also provide
a relevant fraction of the black hole mass density due to growth 
by a accretion a much better knowledge of their space density 
and evolutive properties is needed.
In the following we describe some of the efforts made 
by our team towards the search for highly obscured 
AGN among unidentified faint X--ray sources.

\section{Estimating redshifts for unidentified sources} 

\subsection{A statistical approach}

Combining the X--ray and optical spectroscopic information 
for a well defined sample of hard X--ray selected sources from 
deep {\it Chandra} and XMM--{\it Newton} surveys,  
a striking correlation (Fig.~2) between the X--ray to optical flux ratio 
and the hard X--ray luminosity 
has been found for objects classified as obscured (type 2) 
on the basis of the optical spectrum: 
higher luminosity AGN tend to have higher values of their
$f_X/f_{opt}$ ratio (\cite{ff03}).
There is no evidence of a correlation between the same quantities 
if optically unobscured (i.e. broad line type 1 
quasars) sources are considered. 

\begin{figure}
  \includegraphics[height=.65\textheight,angle=-90,bbllx=150pt,bblly=15pt,bburx=520pt,bbury=740pt]{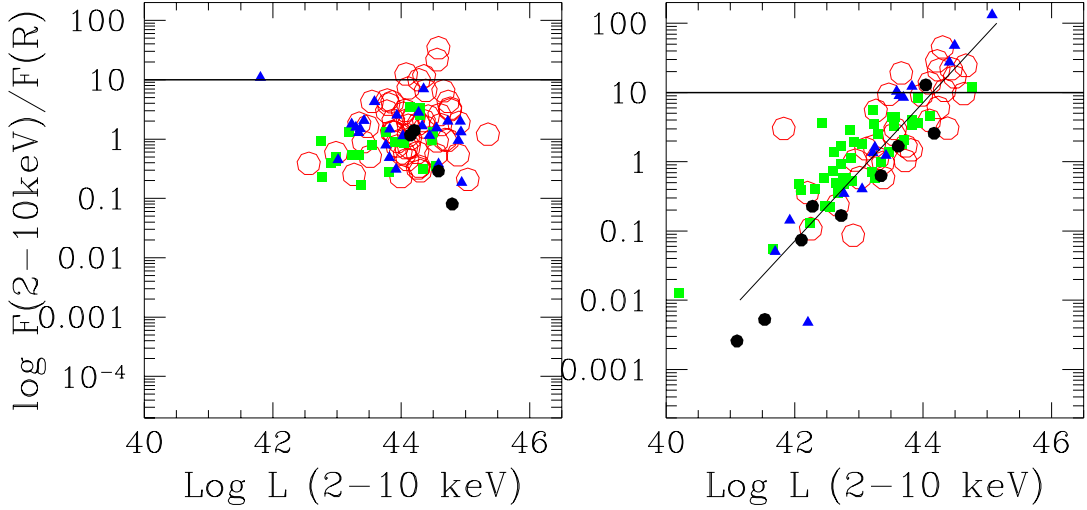}
\caption{The X--ray to optical flux ratio as a function of the 
X--ray luminosity for optically unobscured type 1 AGN (left panel) 
and optically obscured type 2 AGN (right panel). The horizontal line 
corresponds to log$f_X/f_{opt}$=1. The
solid diagonal line in the right panel represents the best linear regression
between log($f_X/f_{opt}$) and logL$_{2-10keV}$. Big open circles are from 
{\tt HELLAS2XMM}, small filled symbols from {\it Chandra} surveys.} 
\end{figure}

Making use of the above described correlation and assuming 
the ratio between type 1 and type 2 objects derived from the 
available identification of high $f_X/f_{opt}$ objects, 
mainly from the {\tt HELLAS2XMM} survey at bright X--ray fluxes,
it is then possible 
to predict luminosities, and therefore redshifts, 
of unidentified faint sources.

The redshift distribution of the sources with a spectroscopically
confirmed redshift shows a sharp decrease at $z >$1.2 which is similar 
to that observed in the {\tt CDFN} (\cite{barg02}) and 
{\tt CDFS} (\cite{has03}). 
When the sources with an estimated redshift are
added, the overall distribution is still peaked 
at $z \simeq$ 1,  
but the decrease above this redshift is less sharp (Fig.~3), thus suggesting
that low redshift peak in the distribution of the sources with
spectroscopic redshift is enhanced by the incompleteness of the
optical identification (in particolar for high $f_X/f_{opt}$ values).

\begin{figure}
  \includegraphics[height=.6\textheight,bbllx=23pt,bblly=140pt,bburx=590pt,bbury=700pt]{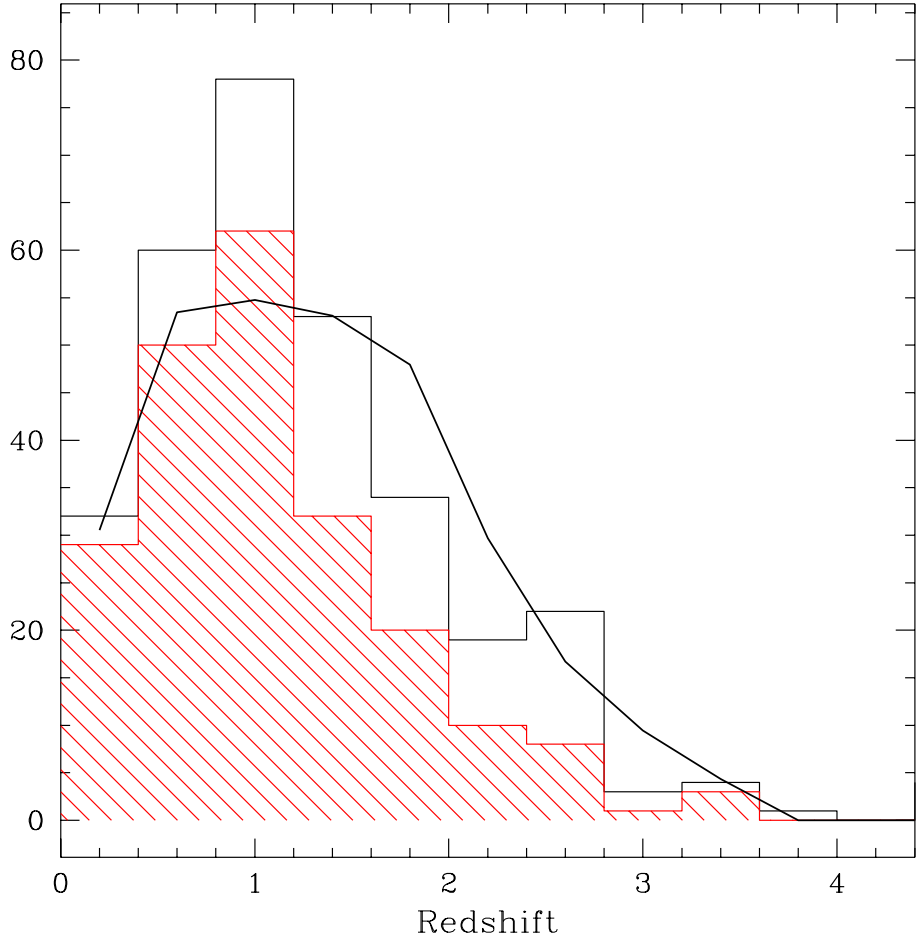}
\caption{The redshift distribution of the sources in the combined 
{\it Chandra} and XMM sample. The shaded historgram represents the 
distribution of spectroscopically identified objects (70\% of the sample),
while the solid histogram the distribution of all the sources computed with 
the method described in the text. The solid line is the prediction of AGN 
synthesis model for the XRB \cite{com01}, folded trough the appropriate 
sky--coverage.}
\end{figure}

\subsection{An approach based on the SED}

An alternative, and more speculative, approach \cite{cbm03}
is based on the assumptions that the
broad band (from near infrared to hard X--ray) spectral energy 
distribution (SED) of sources with high values of their 
$f_X/f_{opt}$ ratios is similar to that of extremely obscured 
Compton thick ($N_H > 1.5 \times$ 10$^{24}$ cm$^{-2}$) 
Seyfert 2 objects in the nearby Universe (see Fig.~4).

\begin{figure}
  \includegraphics[height=.70\textheight,angle=-90,bbllx=150pt,bblly=150pt,bburx=450pt,bbury=730pt]{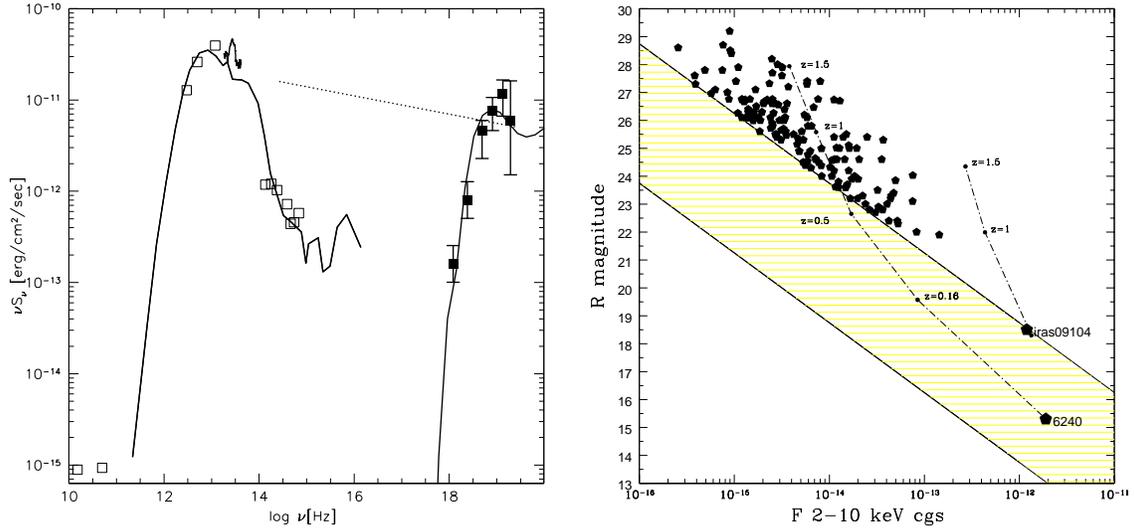}
  \caption{(Left panel): the broad band spectral energy distribution of 
IRAS 09104+4109 (adapted from \cite{fra00}). Right panel: 
The 2--10 keV flux versus the R band magnitude for a sample of 
high $f_X/f_{opt}$ AGN detected in the {\tt HELLAS2XMM} 
survey and in deeper {\it Chandra} and XMM--{\it Newton} 
survey. The dash--dotted lines represent the redshift 
tracks computed as described in the text. Shaded area and diagonal lines
as in Figure 1}. 
\end{figure}

It is relatively straightforward to compute the  
R band optical magnitude 
and the 2--10 keV X--ray flux which would 
be observed for a source with the SED of Fig.~4 (left panel) as a function 
of redshift.
The redshift tracks in the optical magnitude versus X--ray flux plane 
(Fig.~4, right panel) have been normalized to the observed X--ray flux 
and R magnitude of NGC 6240 and IRAS 09104$+$4109 (\cite{vig99}, 
\cite{fra00}).
The two objects are characterized by a similar SED 
but different X--ray luminosities (about $3 \times 10^{44}$ erg s$^{-1}$ 
and 10$^{46}$ erg s$^{-1}$ respectively).
The results clearly indicate, at least at bright X--ray fluxes, 
that the observed high values of
the X--ray to optical flux ratio are consistent with those 
expected by a population of moderately high redshift
($z = 0.5-1.5$), mildly Compton thick ($N_H \simeq$ a few $\times$
10$^{24}$ cm$^{-2}$) AGN with 
X--ray luminosities in the range log$L_X = 44-46$ erg s$^{-1}$.

\subsection{X--ray spectroscopy}

Since its discovery by early X--ray observations, 
the fluorescent FeK$\alpha$ iron emission line at 6.4 keV 
it is by now recognized to be an ubiquitous features in the high 
energy spectra of nearby Seyfert galaxies and an important diagnostic
of the physical and dynamical status of the accreting gas.
The iron line is by far the strongest emission feature 
in the X--ray band and thus can be used as a redshift indicator
if the counting statistic is such to allow a reliable measure
of its centroid energy. A systematic search for redshifted 
iron lines in a flux limited sample of sources 
in the {\tt CDFN} 2 Ms exposure \cite{bau02} indicates a detection 
rate of about 10\%. It is worth to note that 
a much higher percentage is found  
among relatively bright sources with log$f_x/f_{opt} >$ 1 \cite{ccb03}.

\begin{figure}
  \includegraphics[height=.70\textheight,angle=-90,bbllx=140pt,bblly=160pt,bburx=460pt,bbury=730pt]{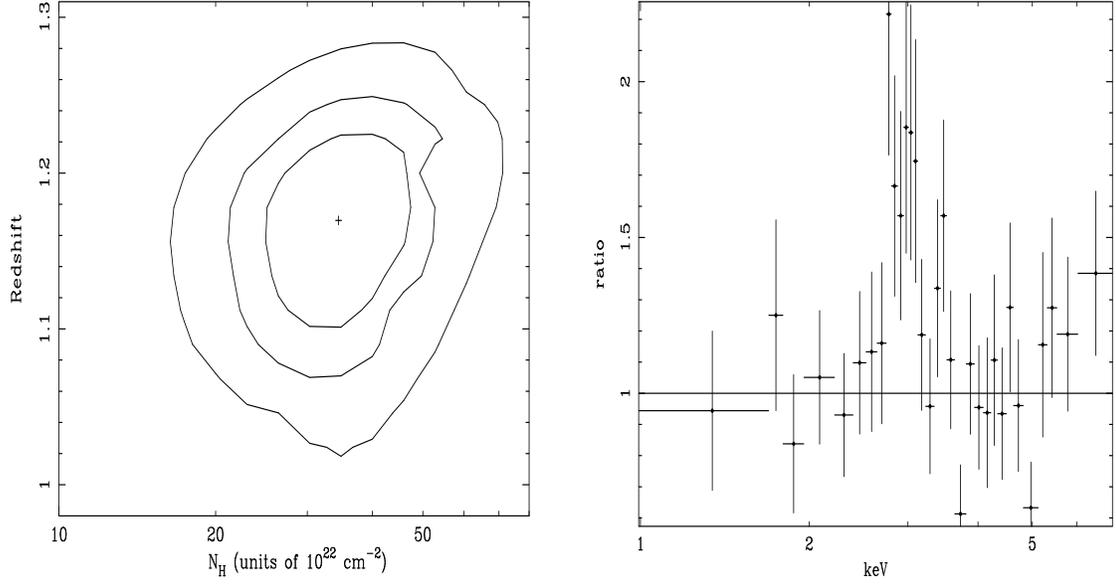}
  \caption{Right panel: the residuals of a power law fit to the {\it Chandra} 
X--ray spectrum of source CXOHDFN 123556.12+621219.1; Left panel: 68, 90 and
99\% confidence contours of the intrinsic column density versus redshift}
\end{figure}

As an example we report here the detection of a 
highly significant emission line feature which is present around
3 keV in the spectrum of the unidentified 
{\it Chandra} source CXOHDFN 123556.12+621219.1
($F_{2-8 keV} = 1.9 \times 10^{-14}$ erg cm$^{-2}$ s$^{-1}$, $R$=23.7,
 $f_x/f_{opt} \simeq$ 18; \cite{davo03}).
The line energy corresponds to an X--ray redshift 
of $z$=1.14$^{+0.11}_{-0.04}$ assuming neutral iron. 
There is also evidence of intrinsic absorption with a column
density $N_H \simeq  3 \times 10^{23}$ cm$^{-2}$.
Given its unabsorbed X--ray luminosity of about 10$^{44}$ erg s$^{-1}$ 
in the 2--10 keV band, this object could be 
classified as an highly obscured, highly luminous type 2 quasar.
The search for iron emission line features will greatly
benefit from the large collecting area of XMM--{\it Newton}
and more X--ray resdhift determination are expected from 
deep XMM--{\it Newton} surveys.

\section{The black hole mass density}

Since the XRB spectral intensity records 
the bulk of the (obscured) accretion power it is possible to compute 
the local density in black holes $\rho_{\bullet}$ due to growth 
by accretion \cite{fi99} using an argument originally proposed by 
Soltan \cite{sol82}:

\begin{equation}
\rho_{\bullet} = 
{k_{bol} \over \eta c^2} (1 + \langle z \rangle)  {4 \pi I_0 \over c} 
\end{equation}

where $I_0$ is the observed intensity of the XRB, $\langle z \rangle$ 
the average redshift of the sources responsible of the bulk of the 
XRB emission,    
$\eta$ the efficiency of accretion and $k_{bol}$ the bolometric correction
from the observed band to the total luminosity. 

A more detailed calculation could be performed if 
the luminosity function $\phi(L)$ and redshift distribution of the sources 
is known:

\begin{equation}
\rho_{\bullet} = {k_{bol} \over \eta c^2} \int {dt \over dz} dz 
\int L \phi(L) dL 
\end{equation}

The value of $\rho_{\bullet}$ can then be compared with that measured from 
local galaxies using the $M_{\bullet}-\sigma$ relation (
\cite{fm00},\cite{geb00}).
A summary of the results of various determination of the
black hole mass density has been recently discussed by \cite{fab03}
and is reported in Table 1.


\begin{table}[t]
\begin{tabular}{l c}
\multicolumn{2}{c}{\bf \normalsize Summary of Mass Densities in
Supermassive Black Holes}\\  
\hline\hline \multicolumn{1}{c}{Method} 
&\multicolumn{1}{c}{$\rho_{\bullet}$ ($10^5$ M$_{\odot}$ Mpc$^{-3}$)}\\
\hline   
XRB spectrum -- pre--Chandra surveys \cite{fi99,erz02} & $6-17$ \\   
XRB spectrum -- new z-distr from Chandra and XMM \cite{fab03} & $\sim4$ \\
Hard X-ray LF \cite{ff03} & 4-6 \\
Hard X-ray LF \cite{cowie03} & $\sim 2$ \\
Bright QSO (from XLF) \cite{yt02} & $\sim 2$ \\
\hline
Local Quiescent Galaxies, $z < 0.0003$ \cite{fer02} & $4-5$\\ \hline\hline
\caption{Adapted from \cite{fer02}}
\end{tabular}
\label{tab:a}
\end{table}
\normalsize
                                               
Although the various estimates agree each other within a factor 2--3,
it has been pointed out \cite{fab03} that taking some of the values at the 
face value could have important astrophysical consequences.

For example if values of the order of 2 $\times$ 10$^5$ M$_{\odot}$ Mpc$^{-3}$
hold, then there would be little room for obscured accretion which 
is in contrast with the X--ray observations previously described, unless 
most of accreting black holes are very efficient ($\eta >$ 0.15, 
\cite{erz02}).
Higher values (4--6 $\times$ 10$^5$ M$_{\odot}$ Mpc$^{-3}$)
of $\rho_{\bullet}$, such as those obtained using the results of
{\it Chandra} and XMM--{\it Newton} surveys \cite{ff03,fi99} and 
a ``standard'' efficiency $\eta \simeq$ 0.1, 
would be consistent with the estimates based on the
$M_{\bullet}$--$\sigma$ relation \cite{fer02}.

In order to obtain robust estimates of the 
black hole mass density due to growth by accretion, 
a much better knowledge of the redshift and luminosity 
distribution (the latter being related to $k_{bol}$) 
of highly obscured AGN is of paramount importance.

\section{Conclusions}

The overall picture emerging from X--ray observations
favours a late formation of the XRB which appears to be dominated
by low luminosity Seyfert galaxies peaking at relatively low redshifts.
However, the physical and evolutionary properties of the obscured AGN 
population responsible for the extragalactic background light 
is still far to be understood. In particular there are evidences 
of a sizeable population of high redshift X--ray luminous quasars
among the still unidentified sources. 
Obscured accretion, which  
can be best investigated in the hard X--ray band, is a fundamental 
tool to understand the evolution of supermassive black holes.
Complete samples of obscured AGN over a range of redshifts
are required. 
To this purpose large area (of the order of a few square degrees) 
hard X--ray surveys provide a promising opportunity.


\begin{theacknowledgments}

I thank the {\tt HELLAS2XMM} team for the extremely good cooperation 
and the permission to use some data in advance of publication and in
particular Marcella Brusa and Cristian Vignali for useful discussions.
I would also thank Joan Centrella  for organizing such 
an interesting meeting and USRA for financial support.
This research is partly supported by the 
Italian Space Agency (ASI) trough I/R/113/01 and I/R/073/01 grants.

\end{theacknowledgments}


\bibliographystyle{aipprocl} 

\end{document}